\documentclass[%
 reprint,
 amsmath,amssymb,
 aps,
]{revtex4-1}

\usepackage{graphicx}
\usepackage{braket}
\usepackage{float}
\usepackage{amsmath}
\usepackage{amsthm}
\usepackage{amsfonts}
\usepackage{amssymb}
\usepackage{mathtools}
\usepackage{calrsfs}
\usepackage{dcolumn}
\usepackage{bm}

\begin{document}

\title{Continuous-Time Quantum Walk on Penrose Lattice}
\author{Yimeng Min}
\email{yimengmin@smail.nju.edu.cn}
\affiliation{
Kuang Yaming Honors School, Nanjing University, Nanjing, Jiangsu 210093, China
}
\author{Kun Wang}
\affiliation{Department of Computer Science and Technology, Nanjing University, Jiangsu 210093, China}


\begin{abstract}
 In this paper, we study the quantum walk on the 2D Penrose Lattice, which is intermediate between periodic and disordered structure.  Quantum walk on Penrose Lattice is less efficient in transport comparing to the regular lattices. By calculating the final remaining probability on the initial nodes and estimating the low bound. Our results show that the broken of translational symmetry induces both the localized states and degeneracy of eigenstates at $E=0$, this two differences from regular lattices influence efficiency of quantum walk.  Also, we observe the transition from inefficient to efficient transport after introducing the near hopping terms, which suggests that we can adjust the "hopping strength" and achieve a "phase transition" progress.
\end{abstract}

\pacs{Valid PACS appear here}
\maketitle

\section{\label{sec1}Introduction}
Quantum walk on a network is a fundamental natural process since the quantum dynamics of any discrete system can be reexpressed and interpreted as a single particle quantum walk, which is capable of performing universal quantum computation~\cite{childs2009universal}. Given the unique symmetry for these 2D systems, the transportation behavior in 2D can be an attractive topic for us to study. Quantum walks on graphs provide an abstract  setting in which to  study  such  transport  properties  independent  of  the  other chemical  and  physical  properties  of a physical substance.  They can thus be used to further the understanding of mechanisms behind such properties. In recent years, two types of quantum walks exist in the literature: the continuous-time quantum walk and discrete-time quantum coined walk.

Continuous-Time Quantum  Walk(CTQW) on simple structures, such as, Cayley tree~\cite{mulken2006efficiency1}, line~\cite{Ashwin2000Quantum}~\cite{Abal2006Erratum}, cycle~\cite{Solenov2005Continuous}~\cite{Sorrentino2006Synchronization} hypercube~\cite{Krovi2005Hitting},  dendrimers~\cite{M2006Coherent} and other regular networks with simple topology have been investigated. Also, for simple underlying geometries the quantum problem is directly related to well-known models in polymer and in solid state physics~\cite{Volta2006Quantum}.  For instance, walks on (one-dimensional) chains are readily treated by a  Bloch ansatz ~\cite{M2005Spacetime},~\cite{Ziman2010PRINCIPLES},~\cite{Kittel2011Introduction},  when periodic boundary conditions (PBCs) are implemented. On the other hand, open boundary conditions (OBCs) appear naturally in polymer physics in relation to the Rouse-model~\cite{Doi2010The}.

Although CTQW received much attention and there has been some work about CTQW on general graphs, many questions about CTQW appear to be quite difficult to answer at the present time.  These quantum walks are analytical solvable and directly related to the well-known problems in solid state physics. Recently, the spacetime structures of CTQW on one-dimensional and two-dimensional lattices with periodic boundary conditions have been studied by Oliver M$\ddot{u}$lken et al. The topology of the lattices they considered is oversimplified, i.e., each node is only connected to its two nearest neighbors and the lattices they investigated have translational symmetry, which equals to regular graphs. For regular graphs with translational symmetrical, the dynamics of the quantum transport is determined by the topology of the network. To this end, it is natural to consider quantum transport on lattices with more connectivity and more complex topology, for example, two dimensional quasicrystals without translational symmetry.
\begin{figure}
\centering
  \includegraphics[width=0.50\textwidth]{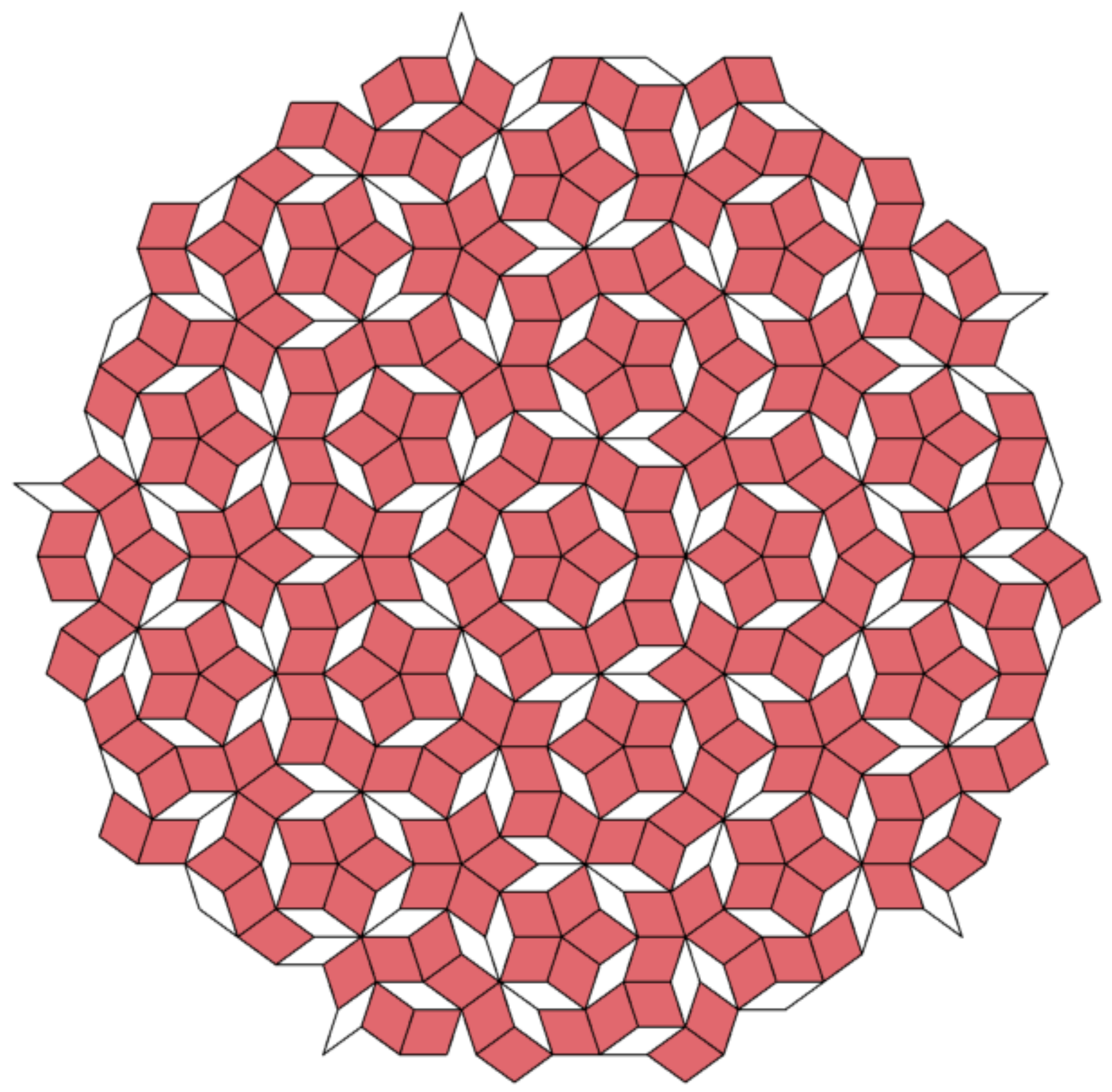}\\
  \caption{Penrose Lattice with "fat rhombus" and "thin rhombus".}\label{fig:7}
\end{figure}

Quasicrystals have well-ordered structures~\cite{PhysRevLett.53.1951} (aperiodic crystals) which fall outside the realm of classical crystallography. For classical lattice with translational symmetry it corresponds to the regular graphs in two-dimensional condition. A new class of materials without translational symmetry was first discovered in Al-Mn alloy in 1984 by D. S. Schechtman~\cite{levine1984quasicrystals}. These aperiodic structures have led to much interdisciplinary activities and many theoretical studies have investigated the properties of quasicrystal models~\cite{wang1987two}~\cite{godreche1986magnetic}~\cite{onoda1988growing}~\cite{stinchcombe1976renormalization}.

In this paper, we systematically investigated the CTQW on Penrose Lattice and explicitly explored the influence of strict localized states on the final distribution of probability. We then investigated the transport efficiency and our results showed the highly degenerate eigenstates and localized states caused by the broken of translational symmetry contribute to the weaken of quantum acceleration effect. By introducing the near hopping terms, the efficiency can be enhanced or weakened by choosing different hopping terms.

The paper is structured as follows:  In the next section we recall the general properties of CTQW on networks. In Sec.~\ref{sec3}, we find that both the localized states and high degeneracy at $E=0$ induced by the rotational symmetry contribute to the inefficient transport. Section.~\ref{sec4} presents the influence of near hopping term on the transport. Conclusions and discussions are given in the last part, Sec.~\ref{sec5}.
\section{\label{sec2}Definition}
The hamiltonian of the particle's quantum walk can be descirbed the adjacency matrix of the graph\cite{Manouchehri2013Physical}.
The tight binding hamiltonian is~\cite{kohmoto1986electronic}:
\begin{equation}\label{eq:ham}
    H=\sum_i |i\rangle \varepsilon_i \langle i | +\sum_{i,j} |i\rangle t_{ij} \langle j |
\end{equation}
where $|i\rangle$ denotes a Wannier state localized at node $i$, and $\varepsilon_i$  is on-site energy.
In our simulation, the elements of adjacency matrix are equal to the hopping integrals $t_{ij}$.
For Penrose Lattice, the number of nodes is N, labeled as $|1\rangle$ to $|N\rangle$, the transition amplitude  $\alpha_{k,j}(t)$ from initial state $|j\rangle$ to state $|k \rangle$ at time t reads  $\alpha_{k,j}(t)=\langle k|U(t)|j\rangle$, where $U(t)=exp(-iHt)$ is the quantum mechanical evolution operator and $H$ is the hamiltonian of the network. The quantum mechanical transition probability is:
\begin{equation}\label{eq:13}
    \pi_{k,j}(t)\equiv|\alpha_{k,j}|^2=|\sum_n \langle k |e^{-i E_n t}|\psi_n\rangle\langle\psi_n|j\rangle|^2
\end{equation}
the long time average (LTA) is\cite{muelken2007inefficient}:
  \begin{equation}\label{eq:14}
 \chi_{k,j}\equiv \lim_{T\rightarrow \infty} \frac{1}{T} \int_0^{T} dt \pi_{k,j}(t)
  \end{equation}
  use Cauchy-Schwarz Inequality:
    \begin{equation}\label{eq:16}
    \chi_{k,j}\geq \lim_{T\rightarrow \infty} |\frac{1}{T} \int_0^{T} dt \sum_n \langle k |e^{-i E_n t}|\psi_n\rangle\langle\psi_n|j\rangle |^2
  \end{equation}
In the limit of $T\rightarrow \infty$, only the states in the subspace $E_n=0$ survive. The return probability $R_j$, which corresponds to the probability on the initial node $\ket{j}$ in the long time limit, reads:
    \begin{equation}\label{eq:17}
R_j=\chi_{j,j}\geq \sum_k {|\langle \psi_{0k} | j \rangle|}^4
  \end{equation}
  Where $\psi_{0k}$ is the eigenstate of eigenvalue $E_n=0$. From Eq.\ref{eq:17}, the low bound of the probability on the initial node depends on the whether the initial node is on the localized states in the subspace of $E=0$. To show that the probability on the initial node is not zero, we must prove that $|\langle \psi_{0_k} | j \rangle|^2$ is not zero, in other words, the initial site ought to locate on the sites of $\psi_{0k}$.
\section{\label{sec3}Continuous-Time Quantum Walk on 2D Penrose Lattice}
\subsection{\label{sec3:level1}Localized States and Return Probability on Penrose Lattice}
For the two dimensional quasicrystals, It has been reported  that there exists three kinds of eigenstates: extended, localized and neither localized nor extended (critical)~\cite{kohmoto1986electronic}~\cite{arai1988strictly}. If the initial state consists of some localized eigenstates, it indicates this part of states will localize in some special areas instead of spreading over a long distance no matter how long the system evolutes.
\begin{figure}
\centering
  \includegraphics[width=0.50\textwidth]{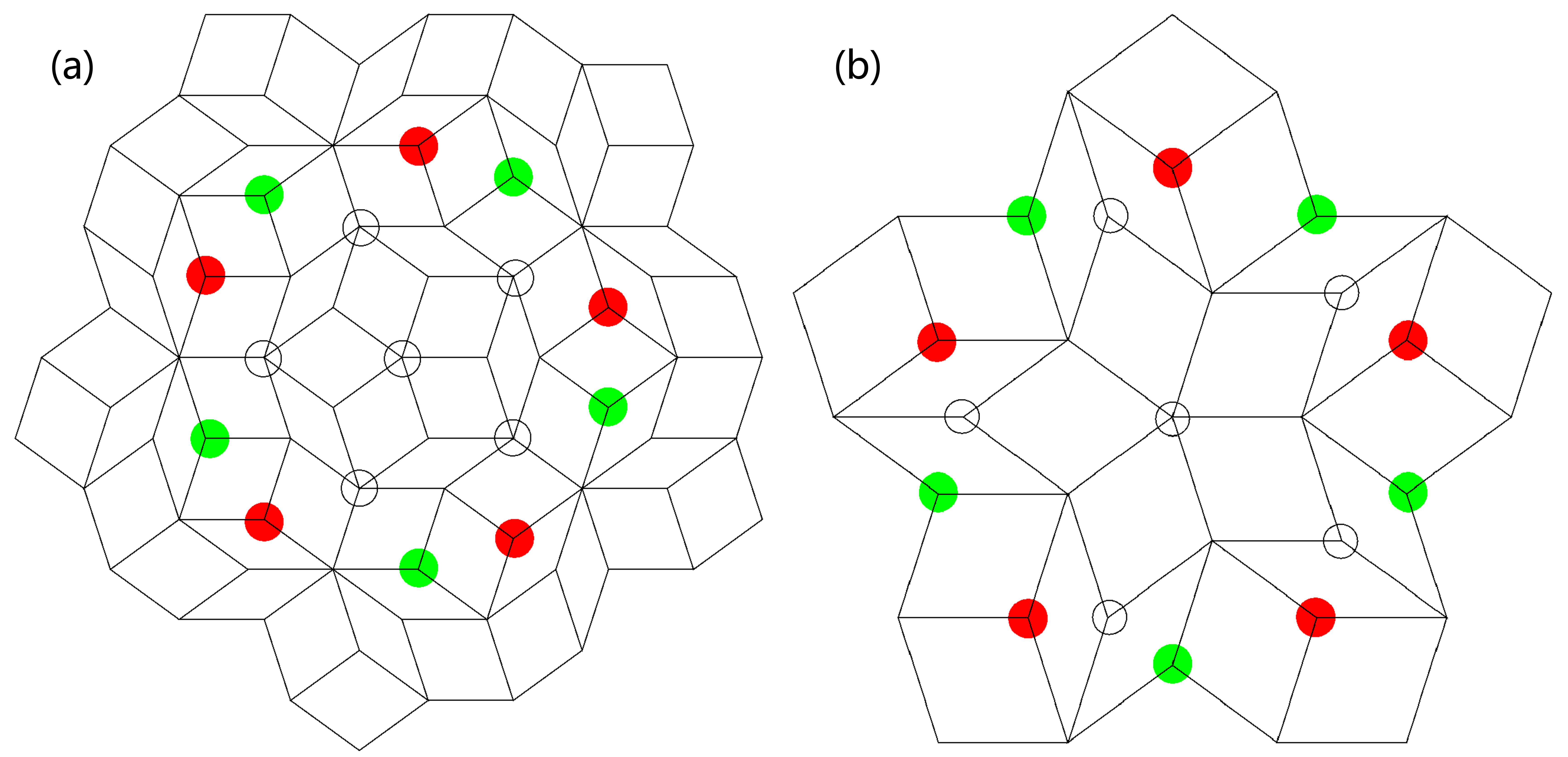}\\
  \caption{The 2 eigenstates found by Kohmoto and Sutherland \cite{kohmoto1986electronic1} with $E=0$. Probability amplitude on the green node is $\frac{1}{\sqrt{10}}$, red node is $-\frac{1}{\sqrt{10}}$ and circle is zero.}\label{fig:10}
\end{figure}
M.Kohmoto~\cite{kohmoto1986electronic} defined a hopping Hamiltonian for independent electrons on a two-dimensional quasiperiodic
Penrose Lattice, where the hopping matrix elements $t_{ij}$ are the same value when the two nodes are connected with a bond. The hopping Hamiltonian simplified the problem and they found some analytical results for the density of states: zero energy eigenstates consist of about $10\%$ of the total number of states and are strictly localized. Then M.Arai\cite{arai1988strictly} continued to study this model and proved that the eigenstates at zero energy are strictly localized and have amplitudes only on some specific nodes, which are most three-degree nodes and some non-three-degree nodes. Fig.~\ref{fig:10} shows two exact eigenstates deduced by Arai and Kohmoto, the probability amplitude on the green node is $\frac{1}{\sqrt{10}}$ and the probability amplitude on the red node is $-\frac{1}{\sqrt{10}}$.
\begin{figure}
\centering
  \includegraphics[width=0.50\textwidth]{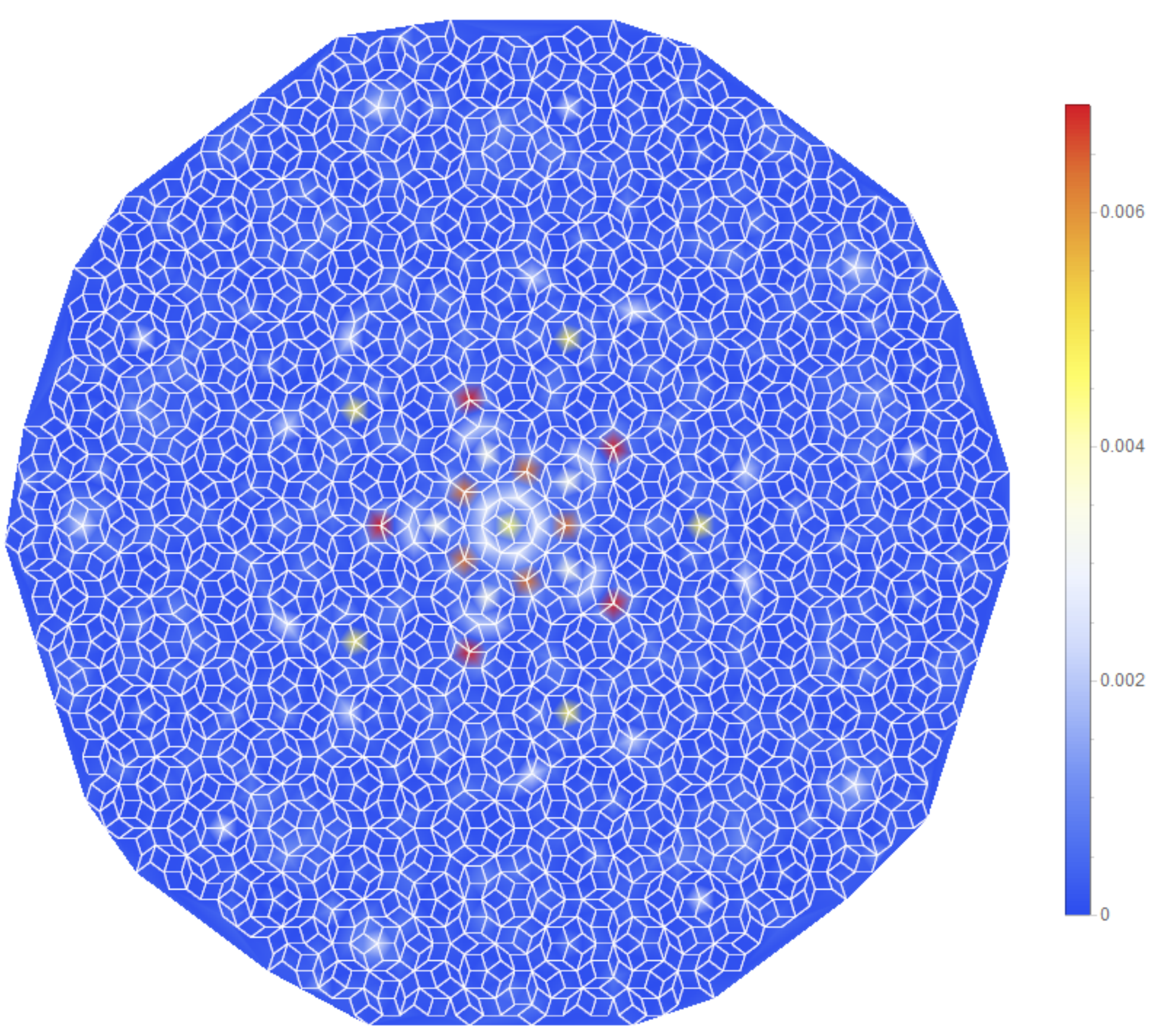}\\
  \caption{CTQW on Penrose Lattice: Initial node: the central node in Fig.\ref{fig:10}(a),time=1000.}\label{fig:13}
\end{figure}
\begin{figure}
\centering
  \includegraphics[width=0.50\textwidth]{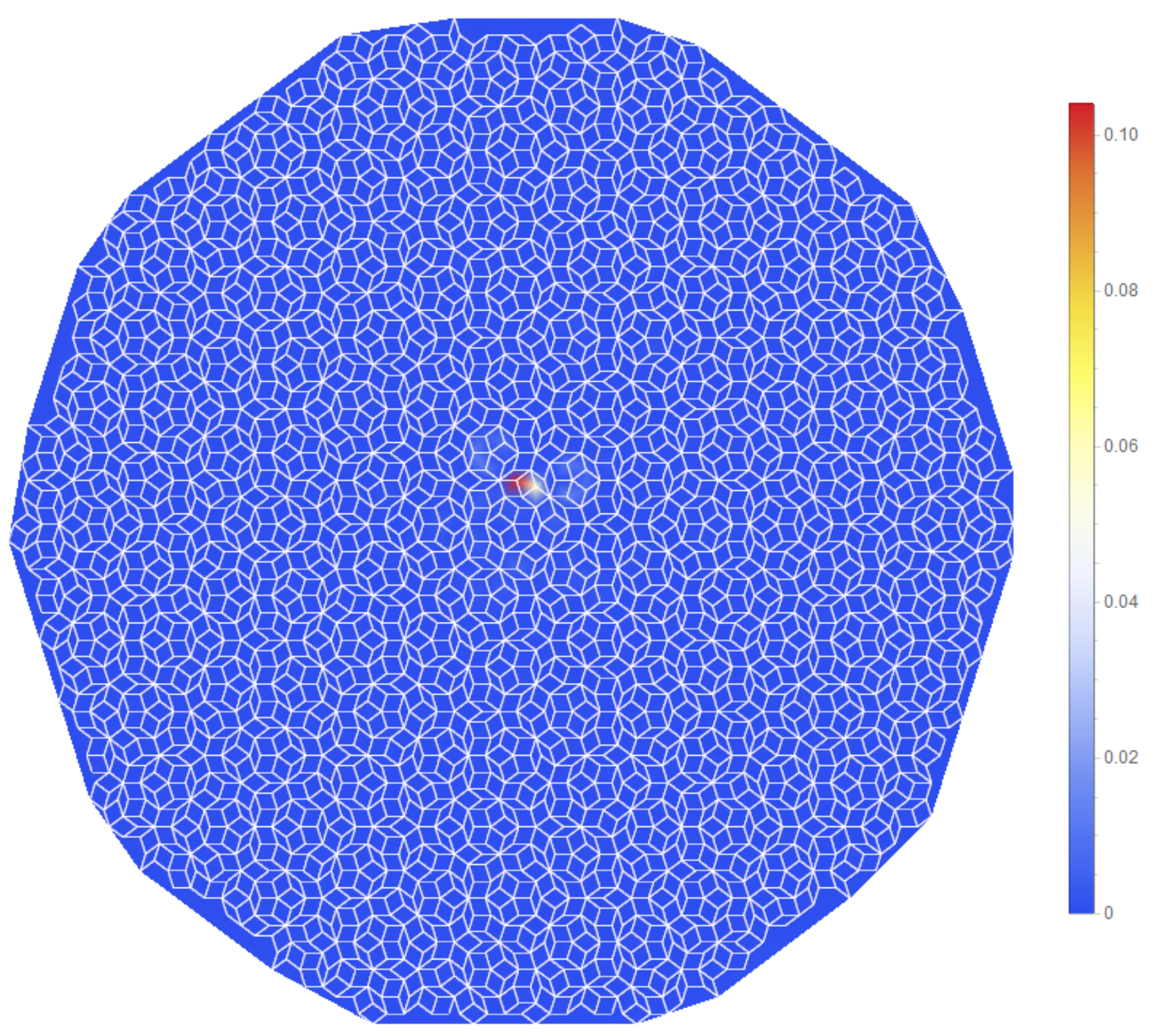}\\
  \caption{CTQW on Penrose Lattice: the green node in Fig.\ref{fig:10}(a),time=1000.}\label{fig:14}
\end{figure}

To verify M.Arai's~\cite{arai1988strictly} conclusion, we choose different initial nodes, the return probability on the 5-degree nodes will be relatively small because most "allowed nodes" are three-degree nodes. Using Eq.~\ref{eq:17} , in the long time limit($T\rightarrow \infty$), $R_j$ will be very small. It indicates that the transport can be considered as efficient because the "particle" will explore other parts away from the network instead of staying on the initial nodes. This efficient transport is common in regular lattices, for regular lattices, the eigenstates are plane waves because of the existence of translational symmetry, the plane waves exist in the whole space and  exhibit periodic oscillation, using Eq.~\ref{eq:17}, the low bound tends to be zero along with the increase of nodes. However, when the translational symmetry is broken, the eigenstates are no longer plane waves any more, if we choose a 3-degree node, which is the red dot in Fig.~\ref{fig:10}(a), the low bound of the return probability will be $\sum_k {|\langle \psi_{0k} | j \rangle|}^4$, which means there will be non-zero probability for the "particle" to stay at the initial node regardless the evolution time and the number of nodes. The results are shown in Fig.~\ref{fig:13} and Fig.~\ref{fig:14}. With the introduction of rotational symmetry, the nodes on the graph are not equal to each other any more, some nodes(3-degree) can be considered as "inefficient nodes" while some nodes(5-degree) can be considered as "efficient nodes", this difference from regular graphs(which correspond to usual lattices with translational symmetry) can be ascribed to the localized states caused by the rotational symmetry.

\begin{figure}
\centering
  \includegraphics[width=0.50\textwidth]{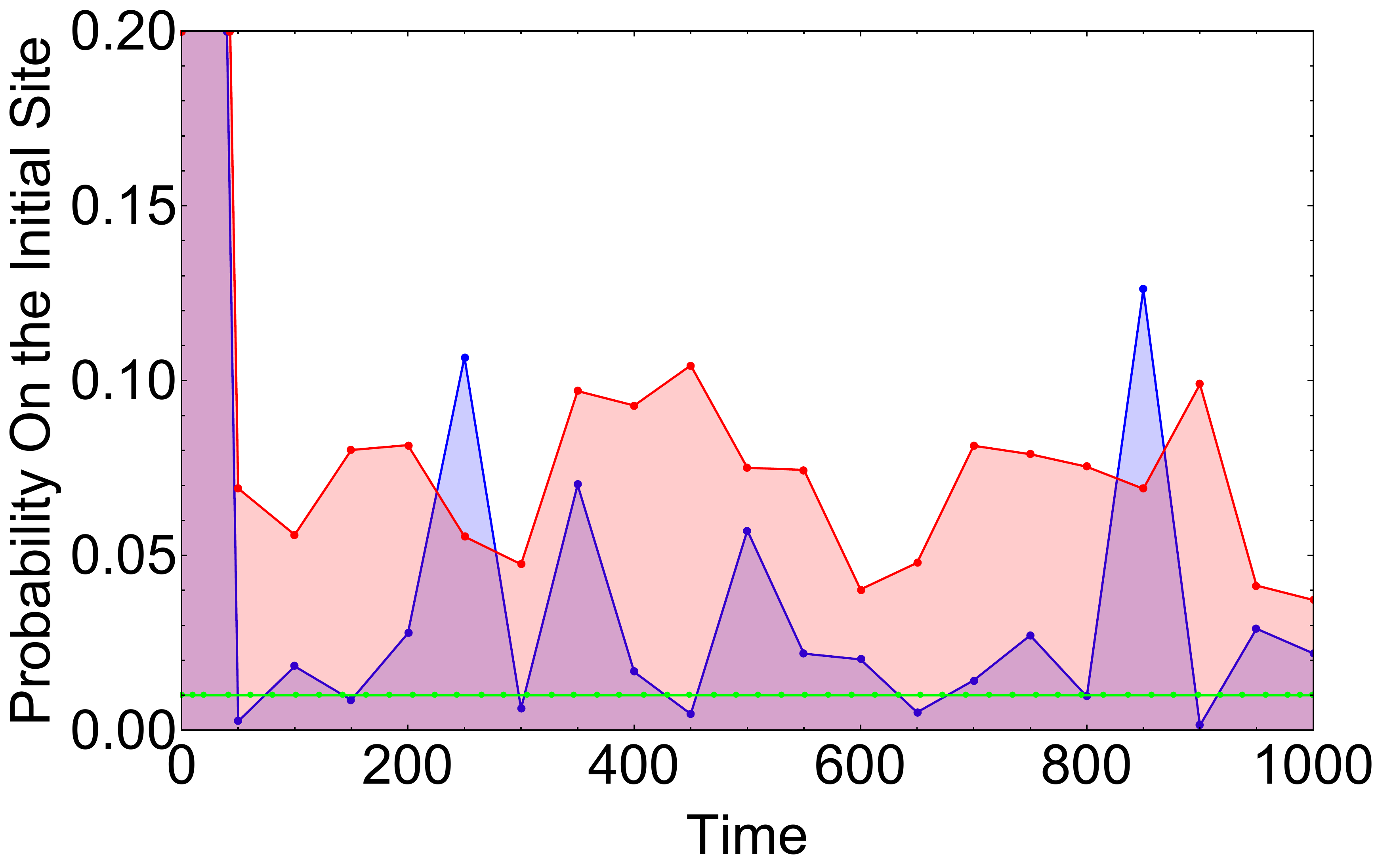}\\
  \caption{Probability with Steps on the different nodes.Blue: the initial node is the central one in Fig.~\ref{fig:10}(a);Red: the initial node is the green one in Fig.\ref{fig:10}(a) }\label{fig:15}
\end{figure}

Fig.~\ref{fig:13} and Fig.~\ref{fig:14} show the probability on every nodes with different initial excitations. In Fig.~\ref{fig:13}, the distribution shows a diffusive trend, the probability decays to very small value in the long time limit. On the other hand, the distribution in Fig.~\ref{fig:14} always oscillates between a certain value. The probability on the initial node is shown in Fig.~\ref{fig:15}. Since most "confined states" locate~\cite{arai1988strictly} on three degree nodes instead of on the five-degree nodes, if we the initial node is the red node in Fig.~\ref{fig:10}(a), the probability $R_j$ in the long time limit will be larger than the condition that the initial node is the central one.
Fig.~\ref{fig:15} indicates that the probability oscillates around a larger value when the initial node is on the confined states, which is the green or red node in Fig.~\ref{fig:10}(a). Considering the long time limit, using Eq.\ref{eq:17} and Fig.\ref{fig:10}, the low bound of $R_j$ should be $|\langle 1 | \frac{1}{\sqrt{10}}\rangle|^4=0.01$, which is the green dashed line in Fig.~\ref{fig:15}.

It is shown that $R_j$ of green/red initial node is relative larger than the expected bound, this may indicate there exists other localized eigenstates who have their probability amplitudes on the green/red node. When the initial node is the central node, $R_j$ can be lower than the green dashed line during the evolution(which equals to be less than 0.01), this suggests that in the subspace $E_0$, the sum  $ \sum_k|\langle \psi_{0k} | central~node \rangle|$ is a relatively small value.
\subsection{\label{sec3:level2}Degeneracy Of States and Inefficient Transport}
Up to now, we have only considered the evolution of single excitation node.
For efficiency of  quantum random walk, the average return probability $\overline{\pi}(t)$ is a good measurement, where $\overline{\pi}(t)=\frac{1}{N} \sum_{j=1}^N \pi_{j,j} (t)$.

\begin{figure}
\centering
  \includegraphics[width=0.50\textwidth]{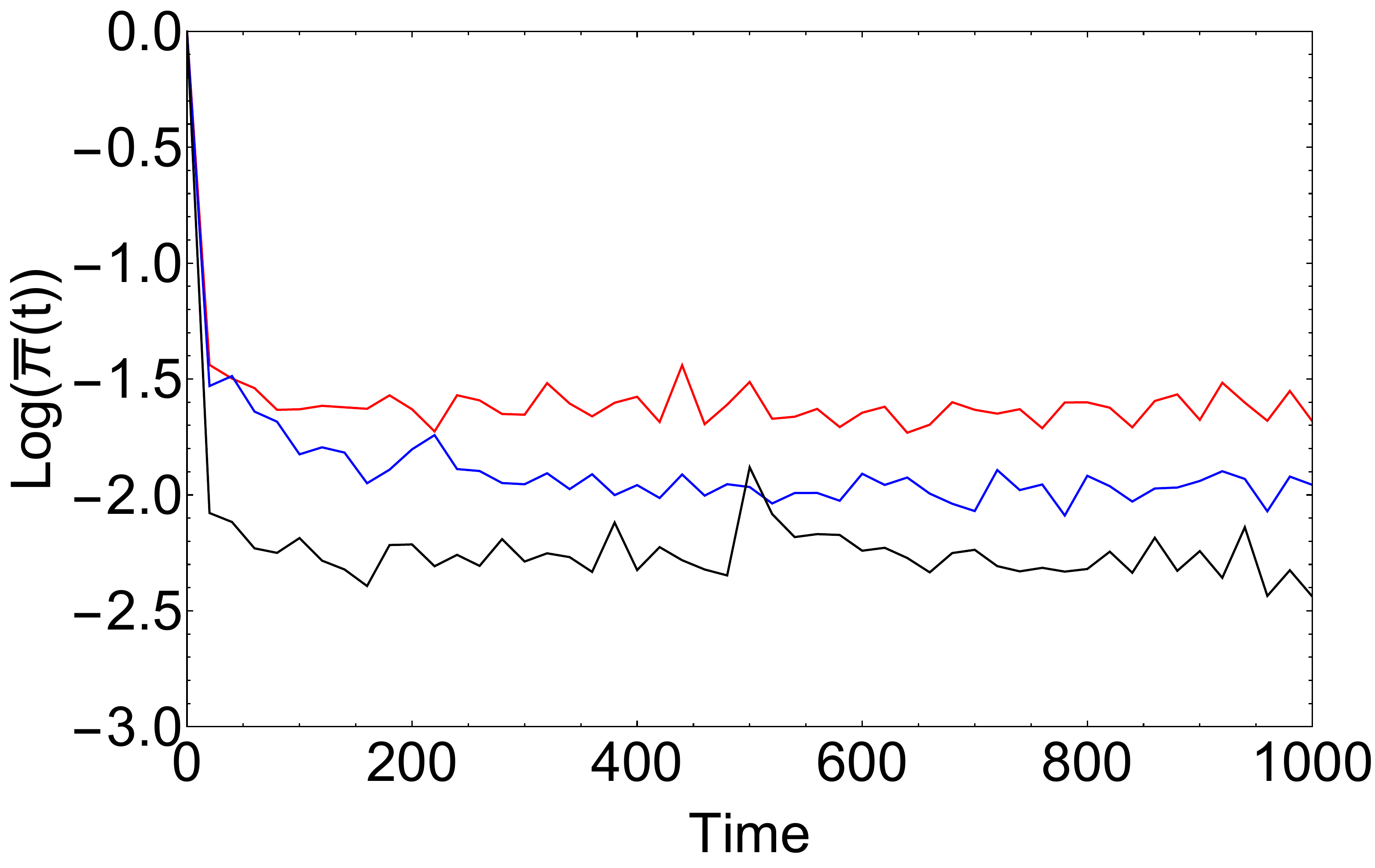}\\
  \caption{The relationship between $\overline{\pi}(t)$ and time with different models.\\
  Black:a=1;b=1.618;c=0.85;\\Red:a=1;b=c=0;\\Blue:a=1;b=1.618;c=0.}\label{fig:den}
\end{figure}
In Ref\cite{muelken2007inefficient}, it has been proved that systems whose eigenvalues have the same degeneracy lead to very efficient transport while networks whose density of states contains few highly degenerate eigenvalues will lead to inefficient performances of quantum walks.
Considering the long time average(LTA) of $ \overline{\pi}(t)$:
\begin{equation}\label{eq:19}
    \overline{\chi}\equiv \lim_{T \rightarrow \infty} \frac{1}{T} \int_0^{\infty} dt \overline{\pi} (t)
\end{equation}
Using Cauchy-Schwarz inequality and we have the low bound:
\begin{equation}\label{eq:20}
   \overline{\pi} (t) \geq |\frac{1}{N} \sum_j \alpha_{j,j}|^2 \equiv |\frac{1}{N} \sum_n e^{-i E_n t}|^2 \equiv |\overline{\alpha}(t)|^2
\end{equation}
Ref~\cite{mulken2006efficiency1} has proved that the envelop of $|\overline{\alpha}(t)|^2$ will decay much quicker than its classical counterpar. As a result, quantum walk can be considered to be more efficient than the corresponding classical random walk. Since $\overline{\pi}(t)$ decays very quickly in time,  the average probability to find the excitation at any node but the initial node increases quickly. As a result, transport on these networks can be considered as efficient. In contrast, if these quantities decay very slowly, we regard the transport on network as being inefficient~\cite{darazs2014transport}.\\
Ref ~\cite{kohmoto1986electronic} has proved that there are about $10\%$ of all eigenstates at $E_n=0$. Assuming there is only one highly degenerate eigenvalue on Penrose Lattice, then for a lattice with $N$ nodes:
\begin{equation}\label{eq:21}
    \overline{\alpha}(t)=\frac{1}{N} [ D_0 e^{-i E_0 t} + \sum_{E_n\neq E_0} D_n e^{-i E_n t} ]
\end{equation}
where $D_0$ is the degenerate degree at $E_0=0$, $D_n$ is the degeneracy of $E_n$($E_n$ is the eigenvalue) and $\frac{D_0}{N}\approx10\%$,
up to order $O(1/N)^2$:
\begin{equation}\label{eq:22}
|\overline{\alpha}(t)|^2\approx \frac{D_0}{N^2}\{D_0+\sum_{E_n\neq E_o} D_n 2 cos[E_n t] \}
\end{equation}
Therefore, for one highly degenerate eigenvalue condition, $\overline{\alpha}(t)$ will fluctuate about ${\frac{D_0}{N}}$ and the low bound $|\overline{\alpha}(t)|^2$ will fluctuate for all times about $(\frac{D_0}{N})^2$.
\begin{figure}
\centering
  \includegraphics[width=0.35\textwidth]{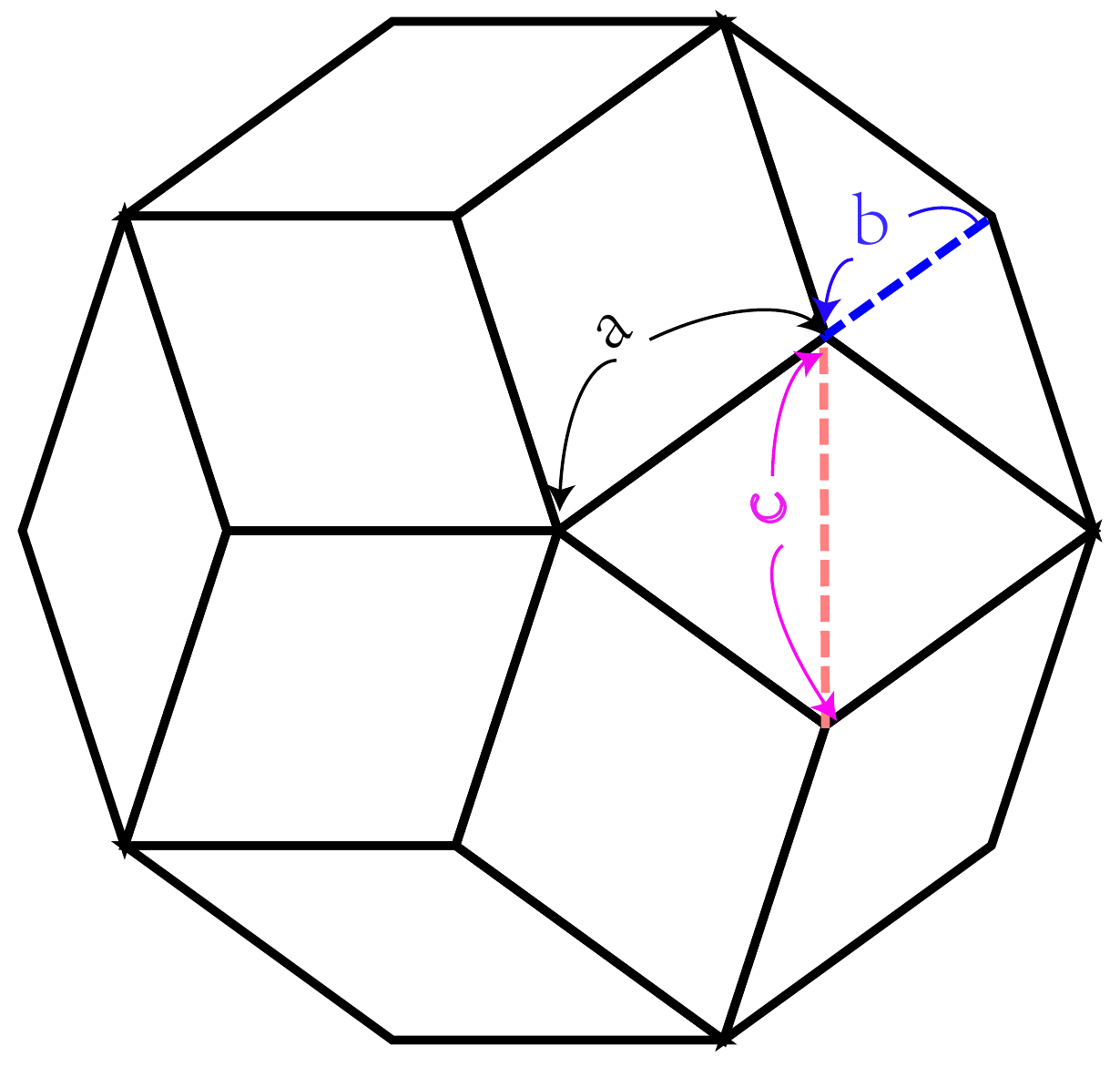}\\
  \caption{Sketch of near hopping term a, b and c}\label{fig:nearhop}
\end{figure}

From the above discuss, the excitation will not explore parts of the Penrose Lattice away from the initial nodes very quickly and the efficiency is not as efficiently as the regular networks.
Considering the degeneracy of the eigenstates of the Penrose Lattice dominates the temporal behaviour, the quantum walk on Penrose Lattice(on average) has larger probability to stay on the initial nodes than quantum walk on regular lattices, which means the quantum walk on Penrose Lattice is inefficient. This difference from regular lattices is caused by the broken of translational symmetry. For a two dimensional lattice with translational symmetry, the eigenstates of these graphs are plane waves and the degeneracy is two for every eigenvalue except $E=0$(whose degeneracy is one). Not only does broken of rotational symmetry break the extended eigenstates, it also induces the non-uniform degree distribution, which results in highly degenerate eigenvalue at $E=0$ and weakens the transportat efficiency of quantum walk.
\section{\label{sec4}Neighborhood Interaction}
\subsection{\label{sec4:level1}Average return probability of universal model}
From Fig.\ref{fig:7}, the distances between nodes have different values. Our previous discussions are based on a simplified hopping model. However, from a more practical view, when the distance between two nodes is close to or even smaller than the edge of the rhombus(EOR), the hopping rate between them should not be zero.  The particle ought to have a certain probability to jump to the other node. The introduction of new hopping terms may influence the transportat efficiency and result in a "phase transition" like Ising model.
\begin{figure}
\centering
  \includegraphics[width=0.50\textwidth]{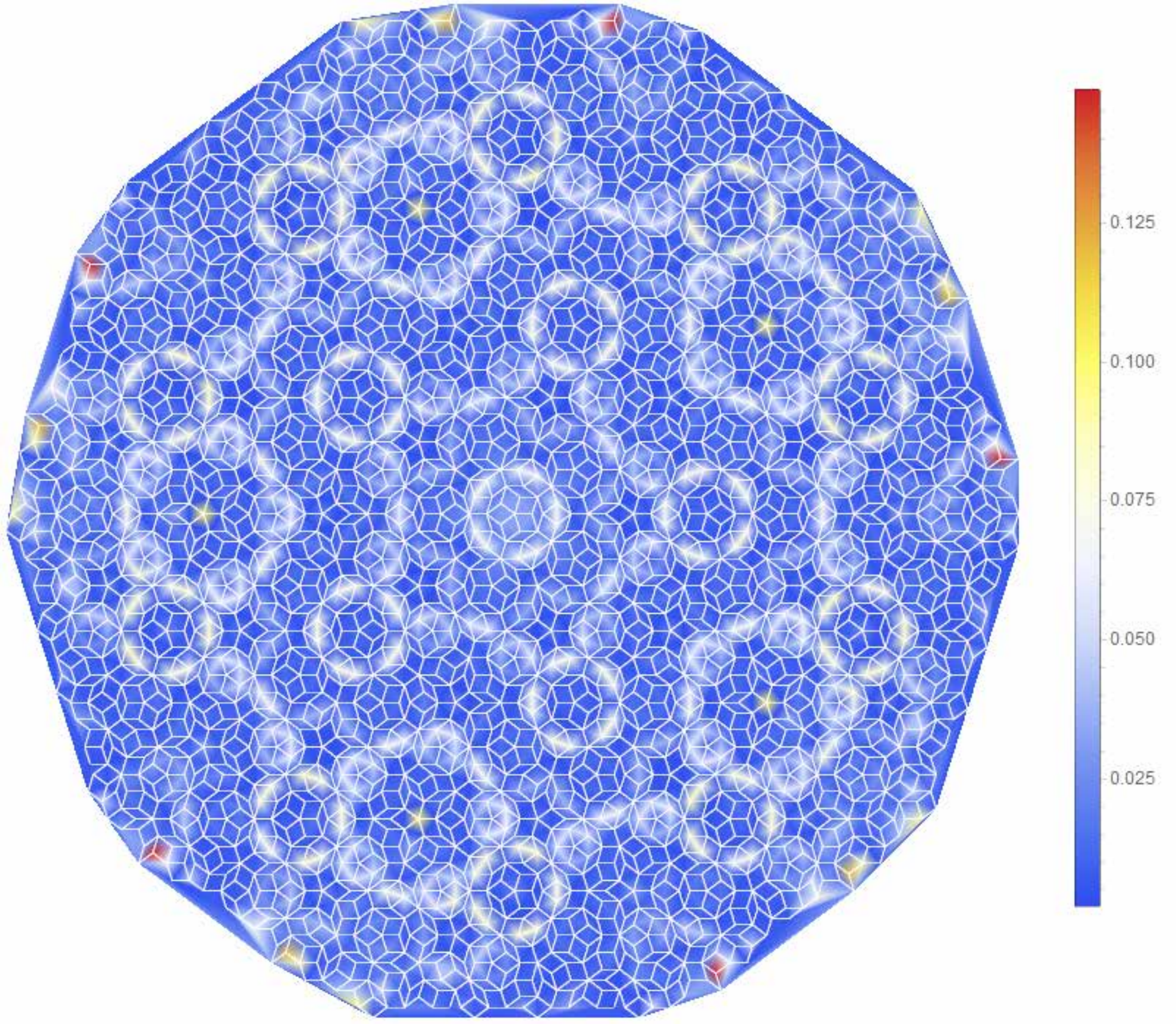}\\
  \caption{$R_j$ of different nodes using Model:a=1;b=c=0;}\label{fig:arbm1}
\end{figure}
In order to be closer to the true physical condition, we define a more universal model:
\begin{itemize}
  \item when the node $i$ and $j$ are on both sides of the rhombus, which means the distance between them is the EOR, $t_{ij}$=a;
  \item when the node $i$ and $j$ are on the short diagonal of the "thin rhombus"(the blue dashed line in Fig.~\ref{fig:nearhop}), which means the distance between them is $2*Sin(18^{\circ})$ of the EOR, $t_{ij}$=b;
  \item  when the node $i$ and $j$ are on the short diagonal of the "fat rhombus"(the red dashed line in Fig.~\ref{fig:nearhop}), which means the distance between them is $2*Sin(36^{\circ})$ EOR, $t_{ij}$=c;
  \item otherwise $t_{ij}=0$.
\end{itemize}
\begin{figure}
\centering
  \includegraphics[width=0.50\textwidth]{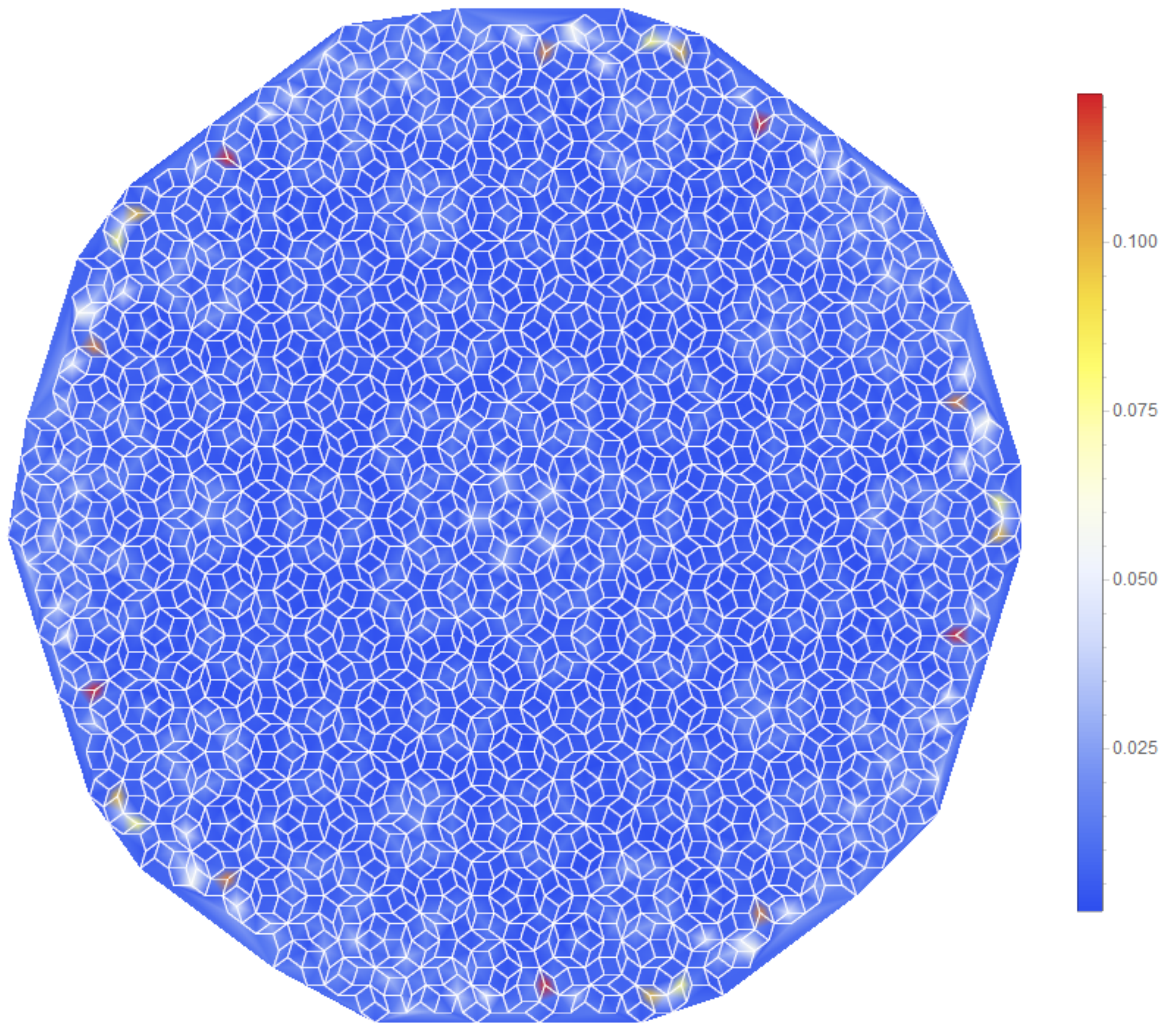}\\
  \caption{$R_j$ of different nodes using Model:a=1;b=1.618;c=0;}\label{fig:arbm2}
\end{figure}
Here we calculate the $\overline{\pi}(t)$  using different models. The results are shown in Fig.\ref{fig:den}, the $\overline{\pi}(t)$ becomes much smaller after we introduce the near hopping terms. When b=$\frac{1}{2*Sin(18^{\circ})}=1.618$, $\overline{\chi}$ decays from 0.026 to 0.016, using Eq.\ref{eq:20} and \ref{eq:22}, the upper bounds of $\frac{D_0}{N}$ are $16.1\%$ and $12.7\%$. After we introduce c=$\frac{1}{2*Sin(36^{\circ})}$, the $\overline{\chi}$ decays sharply to 0.0040, which means the upper bound of $\frac{D_0}{N}$ is $6.3\%$. This suggests the high degeneracy at $E=0$ no longer exists and breakdown of inefficient transport.
\begin{figure}
\centering
  \includegraphics[width=0.50\textwidth]{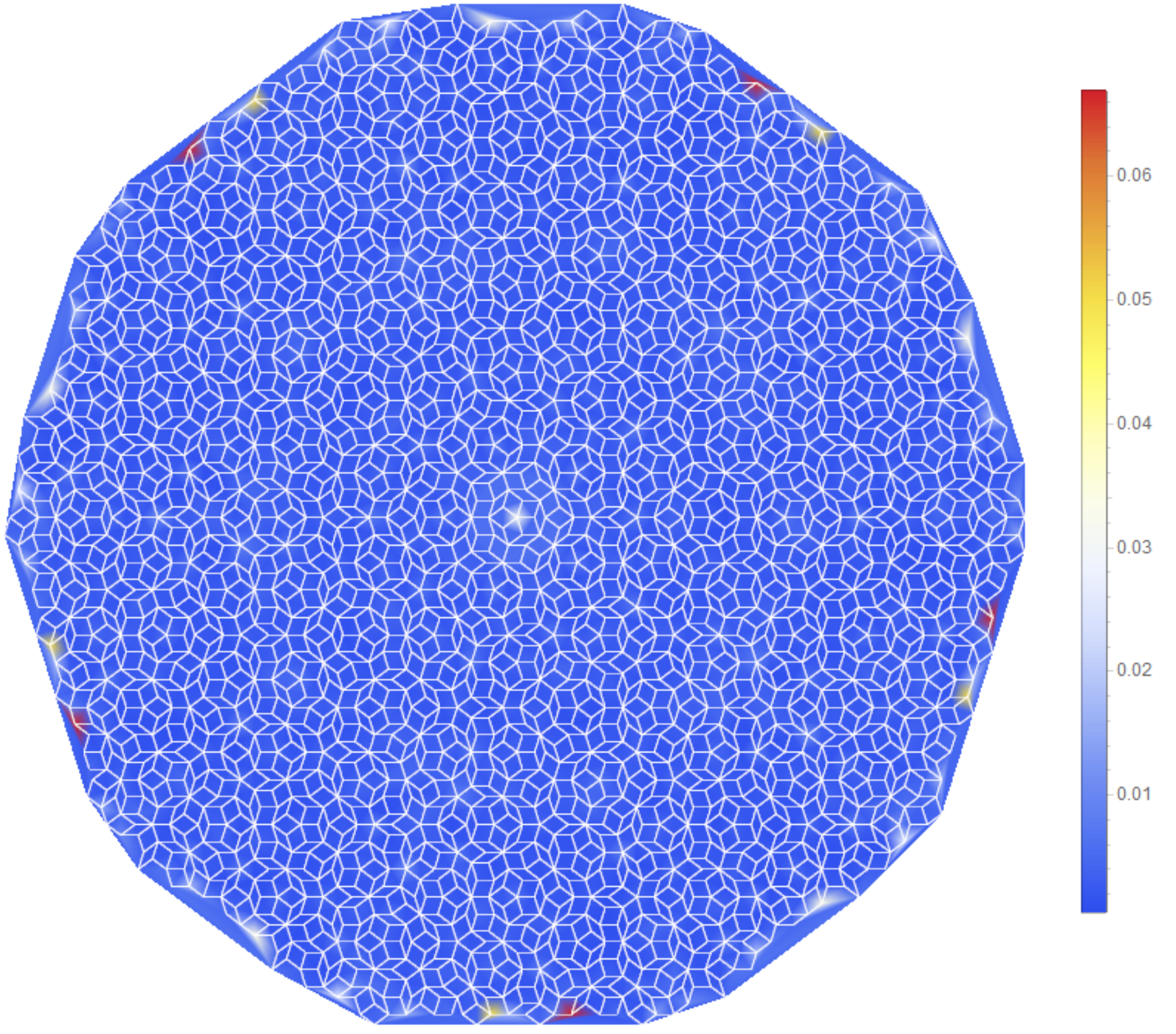}\\
  \caption{$R_j$ of different nodes using Model:a=1;b=1.618;c=0.85;}\label{fig:arbm3}
\end{figure}
\begin{table}
\centering
\begin{tabular}{|c|c|c|c|} \hline
Proportion &\footnotesize{a=1;b=0,c=0}&\footnotesize{a=1;b=1.618;c=0}&\footnotesize{a=1;b=1.618;c=0.85}\\ \hline
$\geq$0.015& $38.6\%$&$13.7\%$& $13.0\%$\\ \hline
$\geq$0.030& $22.0\%$&$2.5\%$& $0.5\%$\\ \hline
$\geq$0.045& $13.8\%$& $1.4\%$ & $0.3\%$ \\ \hline
$\geq$0.060& $8.4\%$& $0.6\%$& $0.2\%$ \\ \hline
$\geq$0.075& $4.0\%$& $0.6\%$& $0\%$\\ \hline
$\geq$0.090& $0.6\%$& $0.4\%$& $0\%$\\ \hline
\end{tabular}
\caption{Proportion of nodes under different hopping parameters}\label{table:1}
\end{table}

From Eq.\ref{eq:20}, since $\overline{\chi}$ decays to a small value and the inefficient transport breaks down, the average return probability ought to show a downtrend. The long time limit of return probability on all nodes using the three models are shown in Fig.~\ref{fig:arbm1},~\ref{fig:arbm2} and ~\ref{fig:arbm3}.
For the condition(a=1,b=0,c=0), as discussed in Sec.\ref{sec3:level1}, due to most strict localized eigenstates locate on the three-degree nodes, these nodes' return probability is relatively high(larger then 0.05). While for other two conditions, after introducing the near hopping term b and c, the return probability becomes smaller than the previous condition. For the model:a=1,b=1.618,c=0, some 7-degree nodes have relative high return probability, which suggests there may exist localized states on these nodes.

The proportions of  different $R_j$ are shown in Table.~\ref{table:1}. It is noticed that there are obvious differences on the proportion of $R_j$ larger than 0.015. For the model a=1,b=0,c=0, the proportion is 38.6$\%$, for the other two models, the value is about 13$\%$. The nodes whose $R_j$ is larger than 0.045 consist about 13$\%$ in model  a=1,b=0,c=0 and are less then 1.5$\%$ in the other two models.

\begin{figure}
\centering
  \includegraphics[width=0.5\textwidth]{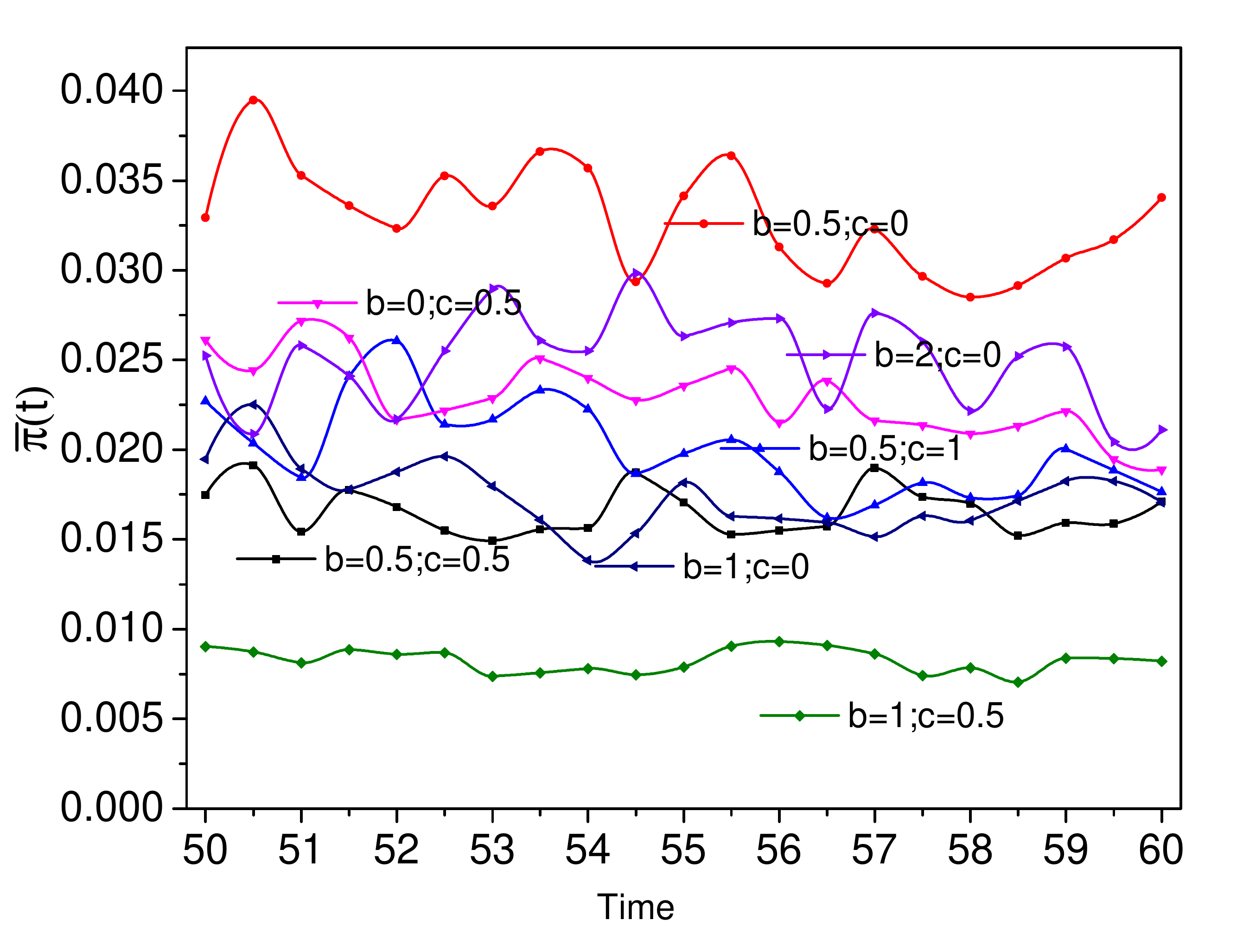}\\
  \caption{Evolution of $\overline{\pi}(t)$ under different hopping parameters(a=1)}\label{fig:evo}
\end{figure}

Does the introduction of new hopping term always make the transport more efficient? From the above discussion, the new hopping terms b and c low the degeneracy at $E=0$ and eliminate the localized states. In Fig.~\ref{fig:evo}, we present the evolution of $\overline{\pi}(t)$ under a relative small time period. It is shown that the model with b=1.0 and c=0.5 oscillates around the smallest value, which indicates the transport under this model can be considered as the most efficient one.

It is also noticed that the efficiency are  non-monotonous with b or c. For example, the LTA of $\overline{\pi}(t)$ using model b=0.5 c=1.0(the blue line in Fig.~\ref{fig:evo}) is larger than the model b=0.5, c= 0.5(the black line in Fig.~\ref{fig:evo}), the increase of hopping parameter c does not make the transport more efficient. Also, when hopping term is a constant(0), with increase of b, $\overline{\pi}(t)$ first decays to a small value and then increases.

As shown in Fig.\ref{fig:arbm2}, over 99$\%$ sites' $R_j$ are less than 0.06 after introducing the near hopping term b; however, more than 13$\%$ nodes' $R_j$ in Fig.\ref{fig:arbm1} are larger than 0.06. The introduction of hopping term c lowers the LTA of return probability further, it decreases and is less than 0.03 for most nodes.
This result suggests the vanish of strict localized eigenstates at $E=0$. Also, it indicates the high degeneracy at $E=0$ no longer exists.
\subsection{\label{sec4:level2}Transition from inefficient to efficient transport}
\begin{figure}
\centering
  \includegraphics[width=0.55\textwidth]{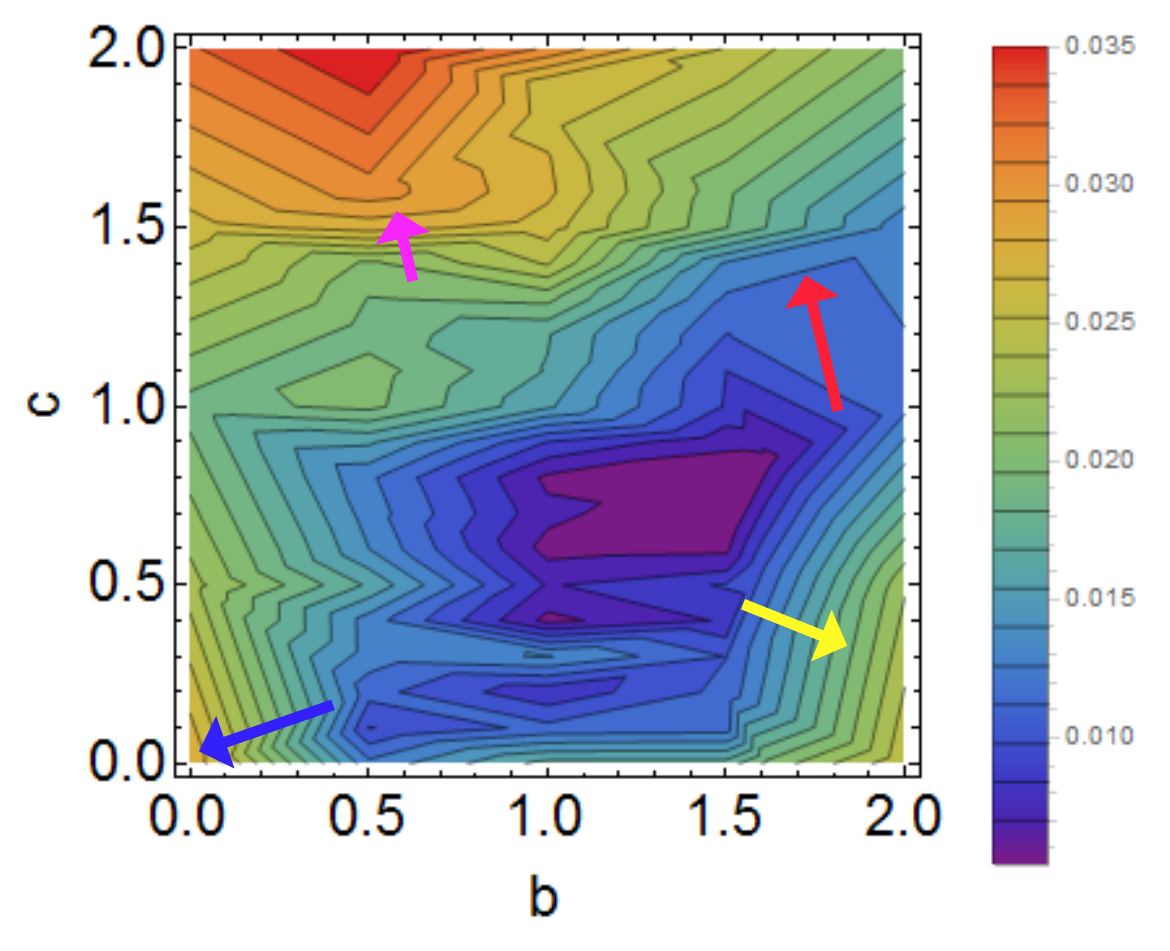}\\
  \caption{$\overline{\chi}$ with different near hopping strength.}\label{fig:ptran}
\end{figure}
From the above discussion,we know that the transport efficiency are  non-monotonous with b or c, so it is interesting for us to find the "most efficient" hopping terms. Generally, a high probability for an excitation to remain at the initial node indicates inefficient transport while the small probability suggests the opposite. From Eq.~\ref{eq:19}, we can characterize a network's transport efficiency by using the $\overline{\chi}$~\cite{Kulvelis2015Universality} . Since the hopping term b and c play fundamental roles in the transport properties, the efficiency can by tuned by adjusting these two parameters. From Fig.\ref{fig:ptran}, the areas of the contour plot are divided into several parts, ranging from less than 0.01 to larger than 0.03, using Eq.~\ref{eq:20}, the degeneracy $\frac{D_0}{N}$ is less than $10\%$ and over $17\%$. When b is 0.5 and c is 2.0, $\overline{\chi}$ is larger than the simplified model a=1, b=c=0, indicating the raise of the $\frac{D_0}{N}$, which is in agreement with the non-monotonicity conclusion. The results show that the inefficient transport on Penrose Lattice can be enhanced or weakened by adjusting the hopping strength to the near nodes.

The contour plot is not isotopic, different areas have obvious differences in the density of the contour lines. From Fig.~\ref{fig:ptran}, the red arrow and the yellow indicate two paths from efficient to  inefficient transport. However, the lengths of this two pathes vary widely. The gradients along the blue and yellow arrow are very steep, there also exists other paths which are sensitive to the hopping strength(the pink arrow), these directions suggest the little variation  can result in an obvious change in the transport property. However, if the hopping terms change along the red arrow, the efficiency changes little and thus can be considered as insensitive.
\section{\label{sec5}Conclusions and discussions}
We generated quantum walk on the 2D Penrose Lattice and found that quantum walk can be considered as inefficient with the absence of the translational symmetry. The extended eigenstates and little degeneracy in regular lattice no longer exist. Both strict localized eigenstates and the high degeneracy at $E=0$ contribute to the inefficient quantum walk. The numerical simulation of the return probability from given excitation is in line with the analytical calculation. Estimations of the degeneracy of eigenstates are also obtained, we calculated a lower bound for the average probability to be still or again at the initial node after some time and the average return probability. The low bound depends on the localized states and high degeneracy and agrees qualitatively well with the exact value. By introducing the near hopping terms and defining the $\overline{\chi}$ to measure the transport properties, we found the efficiency can be enhanced or weakened by adjusting the hopping parameters.
Here we also found there exists some directions which can be considered as more sensitive, when we adjust the hopping parameters along these directions, the efficiency of CTQW varys greatly.
\section{Acknowledgements}
Yimeng want to thank Ping Sheng,Ning Wang and Kwok Yip Szeto in HKUST for the insightful discussions.

\end{document}